\documentclass[twocolumn,showpacs,aps,prd,floatfix,preprintnumbers]{revtex4}
\catcode`\@=12\relax
\usepackage{graphicx}
\usepackage{dcolumn}
\usepackage{epsfig}
\usepackage{psfrag}
\usepackage{amssymb, amsmath}
\usepackage{longtable}
\usepackage{bm}
\usepackage{relsize}
\usepackage{xspace}
\long\def\inst#1{\par\nobreak\kern 4pt\nobreak
    {\it #1}\par\vskip 10pt plus 3pt minus 3pt}
\def\Y#1S{\ensuremath{\Upsilon{(#1S)}}\xspace}
\def\FourS {\Y4S}

\def\babar{\mbox{\slshape B\kern-0.1em{\smaller A}\kern-0.1em
    B\kern-0.1em{\smaller A\kern-0.2em R}}}
\def\pep2{PEP-II}
\def\B       {\ensuremath{B}\xspace}
\def\Bu      {\ensuremath{B^+}\xspace}
\def\Bbar    {\kern 0.18em\overline{\kern -0.18em B}{}\xspace}

\def\ee         {\ensuremath{e^+e^-}\xspace}
\def\mumu       {\ensuremath{\mu^+\mu^-}\xspace}
\def\chicone  {\ensuremath{\chi_{c1}}\xspace}
\def\psitwos  {\ensuremath{\psi{(2S)}}\xspace}
\def\Zp  {\ensuremath{Z(4430)^+}\xspace}

\def\Za  {\ensuremath{Z_1(4050)^+}\xspace}
\def\Zb  {\ensuremath{Z_2(4250)^+}\xspace}
\def\jpsi     {\ensuremath{{J\mskip -3mu/\mskip -2mu\psi\mskip 2mu}}\xspace}
 
\def\Bp      {\ensuremath{\Bu}\xspace}

\def\Bz      {\ensuremath{B^0}\xspace}
\def\Bzb     {\ensuremath{\Bbar^0}\xspace}
\def\mes        {\mbox{$m_{\rm ES}$}\xspace}

\def\DeltaE     {\mbox{$\Delta E$}\xspace}

\def\Kp    {\ensuremath{K^+}\xspace}
\def\Km    {\ensuremath{K^-}\xspace}
\def\piz   {\ensuremath{\pi^0}\xspace}
\def\pip   {\ensuremath{\pi^+}\xspace}
\def\pim   {\ensuremath{\pi^-}\xspace}
\def\KS    {\ensuremath{K^0_{\scriptscriptstyle S}}\xspace}
\def\KL    {\ensuremath{K^0_{\scriptscriptstyle L}}\xspace}
\def\BR         {{\ensuremath{\cal B}\xspace}}
\def\invfb   {\ensuremath{\mbox{\,fb}^{-1}}\xspace}

\def\Bzchikpi {\ensuremath{\Bzb\to\chicone \Km \pip}\xspace}
\def\Bpchikpi {\ensuremath{\Bp\to\chicone \KS \pip}\xspace}

\newcommand{\gevc}{\ensuremath{{\mathrm{\,Ge\kern -0.1em V\!/}c}}\xspace}
\newcommand{\mevc}{\ensuremath{{\mathrm{\,Me\kern -0.1em V\!/}c}}\xspace}
\newcommand{\gevcc}{\ensuremath{{\mathrm{\,Ge\kern -0.1em V\!/}c^2}}\xspace}
\newcommand{\mevcc}{\ensuremath{{\mathrm{\,Me\kern -0.1em V\!/}c^2}}\xspace}
\newcommand{\mev}{\ensuremath{\mathrm{\,Me\kern -0.1em V\!}}\xspace}
\newcommand{\gev}{\ensuremath{\mathrm{\,Ge\kern -0.1em V\!}}\xspace}
\newcommand{\gevcccc}{\ensuremath{{\mathrm{\,Ge\kern -0.1emV^2\!/}c^4}}\xspace}

\begin{document}
\newcommand{\BABARPubYear}    {11}
\newcommand{\BABARPubNumber}  {022}
\newcommand{\SLACPubNumber} {14783}
\preprint{BABAR-PUB-\BABARPubYear/\BABARPubNumber\\}
\preprint{SLAC-PUB-\SLACPubNumber}

\title{\boldmath Search for the \Za and \Zb states in \Bzchikpi and \Bpchikpi}

%
\author{J.~P.~Lees}
\author{V.~Poireau}
\author{V.~Tisserand}
\affiliation{Laboratoire d'Annecy-le-Vieux de Physique des Particules (LAPP), Universit\'e de Savoie, CNRS/IN2P3,  F-74941 Annecy-Le-Vieux, France}
\author{J.~Garra~Tico}
\author{E.~Grauges}
\affiliation{Universitat de Barcelona, Facultat de Fisica, Departament ECM, E-08028 Barcelona, Spain }
\author{M.~Martinelli$^{ab}$}
\author{D.~A.~Milanes$^{a}$}
\author{A.~Palano$^{ab}$ }
\author{M.~Pappagallo$^{ab}$ }
\affiliation{INFN Sezione di Bari$^{a}$; Dipartimento di Fisica, Universit\`a di Bari$^{b}$, I-70126 Bari, Italy }
\author{G.~Eigen}
\author{B.~Stugu}
\affiliation{University of Bergen, Institute of Physics, N-5007 Bergen, Norway }
% ...
\collaboration{The \babar\ Collaboration}
\noaffiliation

\date{\today}

\begin{abstract}

We search for the \Za and \Zb states, reported by the Belle Collaboration, decaying to $\chicone \pip$ in the
decays \Bzchikpi  and $\Bp \to \chicone \KS \pip$ where 
$\chicone \to \jpsi \gamma$. The data were
collected with the \babar\ detector at the SLAC PEP-II
asymmetric-energy $e^+e^-$ collider operating at center-of-mass energy
10.58 \gev\/, and correspond to an integrated luminosity
of 429 fb$^{-1}$. In this analysis, we model the background-subtracted, efficiency-corrected $\chicone\pip$
mass distribution using the $K \pi$ mass distribution and the corresponding normalized $K \pi$ Legendre polynomial 
moments, and then test the need for
the inclusion of resonant structures in the description of the $\chicone \pip$ mass distribution.
No evidence is found for the \Za and \Zb resonances,
and 90\% confidence level upper limits on the branching fractions
are reported  for the corresponding $B$-meson decay modes.

\end{abstract}

\maketitle
\section{Introduction}
The Belle Collaboration has reported the observation of two 
resonance-like structures in the study of \Bzchikpi~\cite{belle3}. These are labeled as \Za and \Zb, both decaying to 
$\chicone \pip$~\cite{PDG}.
The Belle Collaboration also reported 
the observation of a resonance-like structure, $\Zp \to \psitwos\pip$ in the analysis 
of  $\B \to \psitwos K \pi$~\cite{belle1,belle2}. 
These claims have generated a great deal of 
interest~\cite{theory}. Such
states must have a minimum quark content $c\bar c \bar d u$, and
thus would represent an unequivocal manifestation of four-quark meson states.

The \babar \ Collaboration did not see the \Zp in an analysis of the decay $\B \to \psitwos K \pi$~\cite{babarz}.
Points of discussion are:
\begin{itemize}
\item{}
The method of making slices of a three-body \B decay Dalitz plot can produce peaks which may 
be due to interference effects, not resonances.
\item{} The angular structure of the $\B \to \psitwos K \pi$ decay is rather complex and cannot be described adequately by 
only the two variables used in a simple Dalitz plot analysis.
\end{itemize}

In the \babar \ analysis~\cite{babarz}, the $\B \to \jpsi K \pi$ decay does not show evidence for 
resonances neither in $\jpsi\pi$ nor in $\jpsi K$ systems. All resonance activity seems confined 
to the $K \pi$ system. It is also observed that the
angular distributions, expressed in terms of the $K \pi$ Legendre polynomial moments, show strong similarities
between $\B \to \psitwos K \pi$ and $\B \to \jpsi K \pi$ decays. Therefore, the angular information provided by
the $\B \to \jpsi K \pi$ decay can be used to describe the $\B \to \psitwos K \pi$ decay. It is also observed that
a localized structure in the $\psitwos \pi$ mass spectrum would yield high angular momentum  Legendre polynomial 
moments in the $K \pi$ system. Therefore, a good description of the $\psitwos \pi$ data using only $K \pi$ moments up to $L=5$ also suggests the absence of narrow
resonant structure in the $\psitwos \pi$ system.

In this paper, we examine $B \to \chicone K \pi$ decays following an analysis procedure similar to that used in Ref.~\cite{babarz}. 
In contrast to the analysis of Ref~\cite{belle3}, we model the background-subtracted, efficiency-corrected $\chicone\pip$
mass distribution using the $K \pi$ mass distribution and the corresponding normalized $K \pi$ Legendre polynomial 
moments, and then test the need for
the inclusion of resonant structures in the description of the $\chicone \pip$ mass distribution.

This paper is organized as follows. 
A short description of the \babar\ experiment is given in Sec. II, and the data selection is described in Sec. III.
Section IV shows the data, while Sec. V and Sec. VI are devoted to the calculation of the efficiency and the extraction of branching
fraction values, respectively. In Sec. VII we describe the fits to the $K \pi$ mass spectra, and in Sec. VIII we show the
Legendre polynomial moments. In Sec. IX we report the description of the $\chicone \pip$ mass spectra, while
Sec. X is devoted to the calculation of limits on the production of the \Za and \Zb resonances. We summarize our results in Sec. XI. 

\section{The \babar\ experiment}
This analysis is based on a data sample
of 429~$\invfb$ recorded at the
\FourS resonance by the \babar\ detector at the \pep2
asymmetric-energy $e^+e^-$ storage rings. 
The \babar\ detector is
described in detail elsewhere~\cite{babar}. 
Charged particles are detected
and their momenta measured with a combination of a 
cylindrical drift chamber (DCH)
and a silicon vertex tracker (SVT), both operating within the
1.5 T magnetic field of a superconducting solenoid. 
Information from
a ring-imaging Cherenkov detector is combined with specific ionization measurements from the 
SVT and DCH to identify charged kaon and pion candidates. 
Photon energy and position are measured with a CsI(Tl) electromagnetic calorimeter (EMC), which is also used to
identify electrons.
The return yoke of the superconducting coil is instrumented with
resistive plate chambers for the identification of muons. For
the later part of the experiment the barrel-region chambers were replaced by limited streamer tubes~\cite{lst}.

\section{Data Selection}
We reconstruct events in the decay modes~\cite{conj}:
\begin{eqnarray}
\Bzchikpi \, ,  \\
\Bp \to \chicone \KS \pip \, ,
\end{eqnarray}
where $\chicone \to \jpsi \gamma$, and $\jpsi \to \mumu$ or $\jpsi \to \ee$. 

For each candidate, we first reconstruct the \jpsi by
geometrically constraining an identified \ee or \mumu
pair of tracks to a common vertex point and requiring a $\chi^2$ fit probability greater than 0.1\%. 
For $\jpsi \to \ee$ we introduce bremsstrahlung energy-loss recovery.
If an electron-associated photon cluster is found in the
EMC, its three-momentum vector is incorporated into the calculation of $m(\ee)$~\cite{brem}. 
The fit to the \jpsi candidates includes the 
constraint to the nominal \jpsi mass value~\cite{PDG}.

A $\KS$ candidate is formed by geometrically constraining a pair of
oppositely charged tracks to a common vertex ($\chi^2$ fit probability greater than 0.1\%). For the two tracks the pion mass is assumed without
particle-identification requirements. The  $\KS$ fit includes the constraint to the nominal mass value.

The $\jpsi$, $K^{\pm}$, and $\pi^{\pm}$ candidates forming a $B$ meson decay
candidate are geometrically constrained to a common 
vertex and a $\chi^2$ fit probability greater than 0.1\% is required. Particle identification is applied to both $K$ and $\pi$ candidates.
The \KS
flight length with respect to the \Bp vertex must be greater than 0.2 cm.

A study of the scatter diagram $E_{\gamma}$ vs. $m(\jpsi \gamma)$ (not shown) reveals that no \chicone signal 
is kinematically possible for
$E_{\gamma}<190$ MeV. Therefore, we consider only photons with a laboratory energy above this value.
We select the \chicone signal within $\pm 2 \ \sigma_{\chicone}$ of the \chicone mass, where $\sigma_{\chicone}$ and the
\chicone mass are obtained from fits to the $\jpsi \gamma$ mass spectra using a Gaussian function for 
the signal
and a $2^{\rm nd}$-order polynomial for the background,
separated by $B$ and 
\jpsi decay mode. The values of $\sigma_{\chicone}$ range from 14.6 \mevcc to 17.6 \mevcc.

We further define $B$ meson decay candidates using the energy
difference $\DeltaE \equiv E^{\ast}_B-\sqrt{s}/2$ in the center-of-mass
(c.m.) frame and the beam-energy-substituted mass defined as
$\mes \equiv \sqrt{((s/2+\vec{p}_i\cdot\vec{p}_B)/E_i)^2-\vec{p}_B^{\,2}}$,
where ($E_i,\vec{p}_i$) is the initial state \ee four-momentum vector in
the laboratory frame and $\sqrt{s}$ is the c.m. energy. In the above expressions $E^{\ast}_B$
is the $B$ meson candidate energy in the c.m. frame, and $\vec{p}_B$ is its laboratory
frame momentum.
The $B$ decay signal events are selected within $\pm 2.0 \ \sigma_{\mes}$ of the fitted central value,
where the $\sigma_{\mes}$ values are listed in Table~\ref{tab:table1} and are determined by fits of a Gaussian function
plus an ARGUS function~\cite{argus} to the data.
 
\begin{figure}[!htb]
\begin{center}
\includegraphics[width=9.5cm]{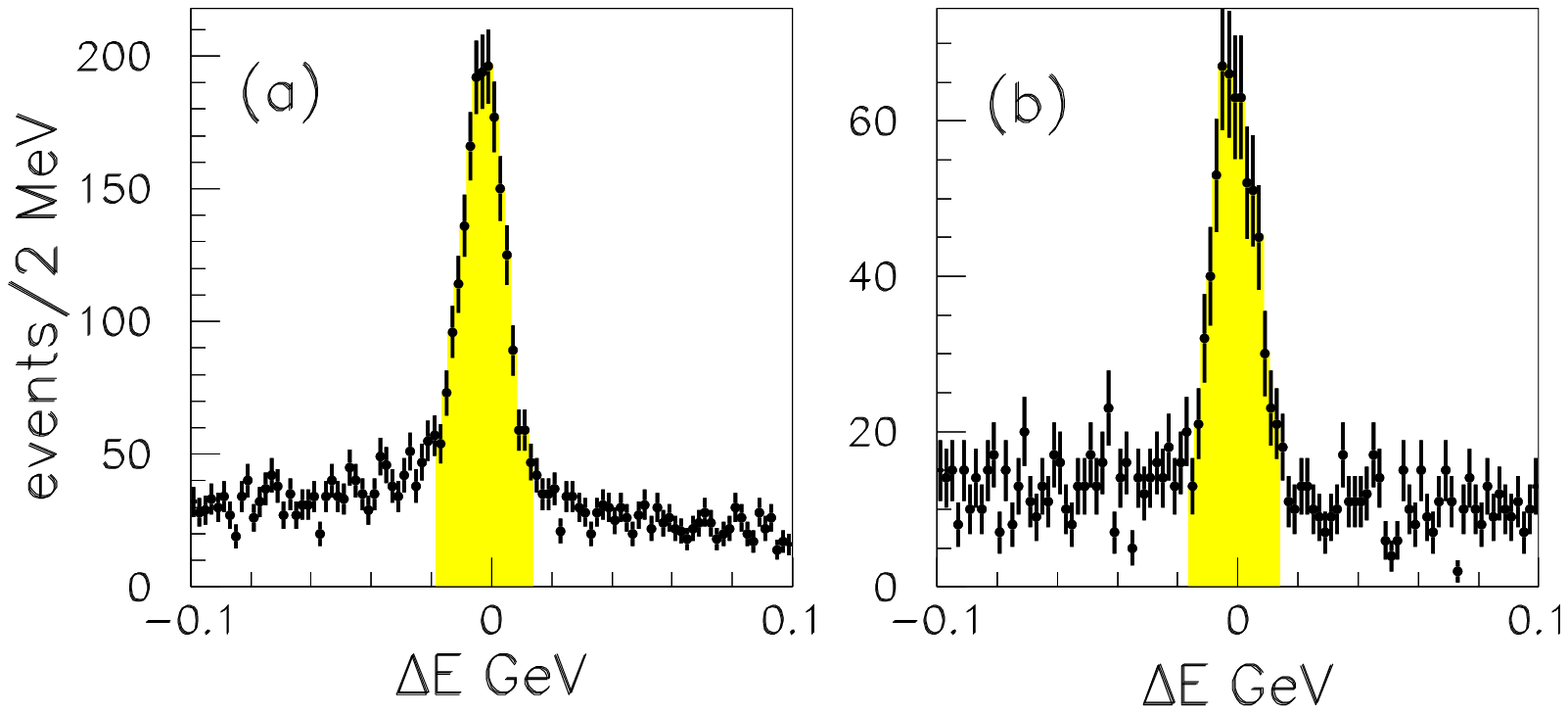}
\caption{Distributions of \DeltaE\ for (a) \Bzchikpi and (b) $\Bp \to \chicone \KS \pip$ summed
over the \jpsi decay modes; the \chicone and \mes selection criteria have been applied. The shaded areas indicate the signal
regions.}
\label{fig:fig1}
\end{center}
\end{figure}

The resulting \DeltaE distributions
have been fitted with a linear background function and a signal Gaussian function whose width values ($\sigma_{\DeltaE}$) are also listed
in Table~\ref{tab:table1}.  Further background rejection is performed by selecting events within $\pm 2.0 \ \sigma_{\DeltaE}$ of zero.
Table~\ref{tab:table1} also gives the values of event yield and purity, where the {\it Purity} is defined as
{\it Signal/(Signal+Background)}.
The \DeltaE distributions shown in Fig.~\ref{fig:fig1} have been summed over the $\jpsi \to \mumu$ and $\jpsi \to \ee$ decay modes. Clear signals of the $B$ decay modes (1) and (2) can be seen.
We obtain 1863 candidates for \Bzchikpi decays with 78\% purity, and  628 $\Bp \to \chicone \KS \pip$ events with 79\% purity.
A study of the \DeltaE and $\jpsi \gamma$ spectra in the sideband regions does not show any $B$ or
\chicone signal respectively. We conclude that the observed background is consistent with being entirely of
combinatorial origin.

\begin{figure}[!htb]
\begin{center}
\includegraphics[width=9.5cm]{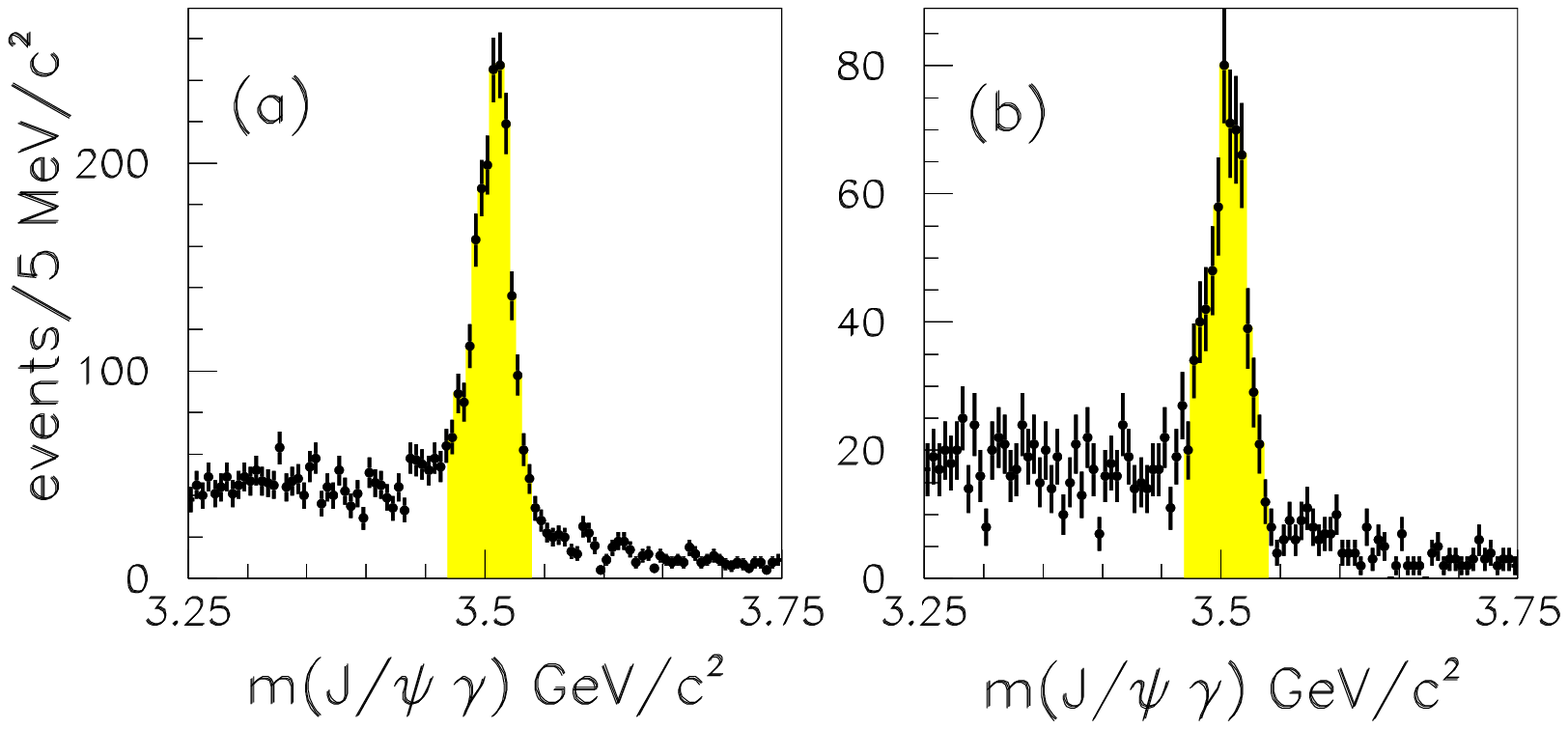}
\caption{The $\jpsi \gamma$ mass distribution for (a) \Bzchikpi and (b) $\Bp \to \chicone \KS \pip$ candidates,
summed over the \jpsi decay modes. The \mes and \DeltaE selection criteria have been applied. The shaded areas indicate the signal
regions.}
\label{fig:fig2}
\end{center}
\end{figure}

The resulting $\jpsi \gamma$ invariant mass distributions for channels (1) and (2) are shown in Fig.~\ref{fig:fig2}. 

In order to estimate the background contribution in the signal region, we define \DeltaE\ sideband
regions in the intervals $(7-9) \ \sigma_{\DeltaE}$ on both sides of zero.
We obtain a ``background-subtracted'' distribution of events by subtracting the corresponding 
distribution for \DeltaE\ sideband events from that of events in the signal region.

\begin{table*}
\caption{Resolution parameter values from fits to the \DeltaE\ and \mes distributions.}
\label{tab:table1}
\begin{center}
\vskip -0.2cm
\begin{tabular}{lcccc}
\hline
Channel & $\sigma_{\DeltaE} \ (\mev)$ & $\sigma_{\mes} \ (\mevcc)$ & events & {\it Purity} \% \cr
\hline \noalign{\vskip4pt}
 \Bzchikpi (\mumu) &6.96 $\pm$ 0.34 & 2.60 $\pm$ 0.10 & 980 & 79.3 $\pm$ 1.3 \cr
 \Bzchikpi (\ee) & 7.81 $\pm$ 0.43 & 2.77 $\pm$ 0.12 & 883 & 77.1 $\pm$ 1.4 \cr
\hline
 $\Bp \to \chicone \KS \pip$ (\mumu) & 6.65 $\pm$ 0.55 & 2.65 $\pm$ 0.27 & 299 & 81.7 $\pm$ 2.2 \cr
 $\Bp \to \chicone \KS \pip$ (\ee) & 7.52 $\pm$ 0.70 &  2.65 $\pm$ 0.18 & 329 & 77.5 $\pm$ 2.3 \cr
\hline
\end{tabular}
\end{center}
\end{table*}

\section{Dalitz plots}

\begin{figure*}
\begin{center}
\includegraphics[width=6cm]{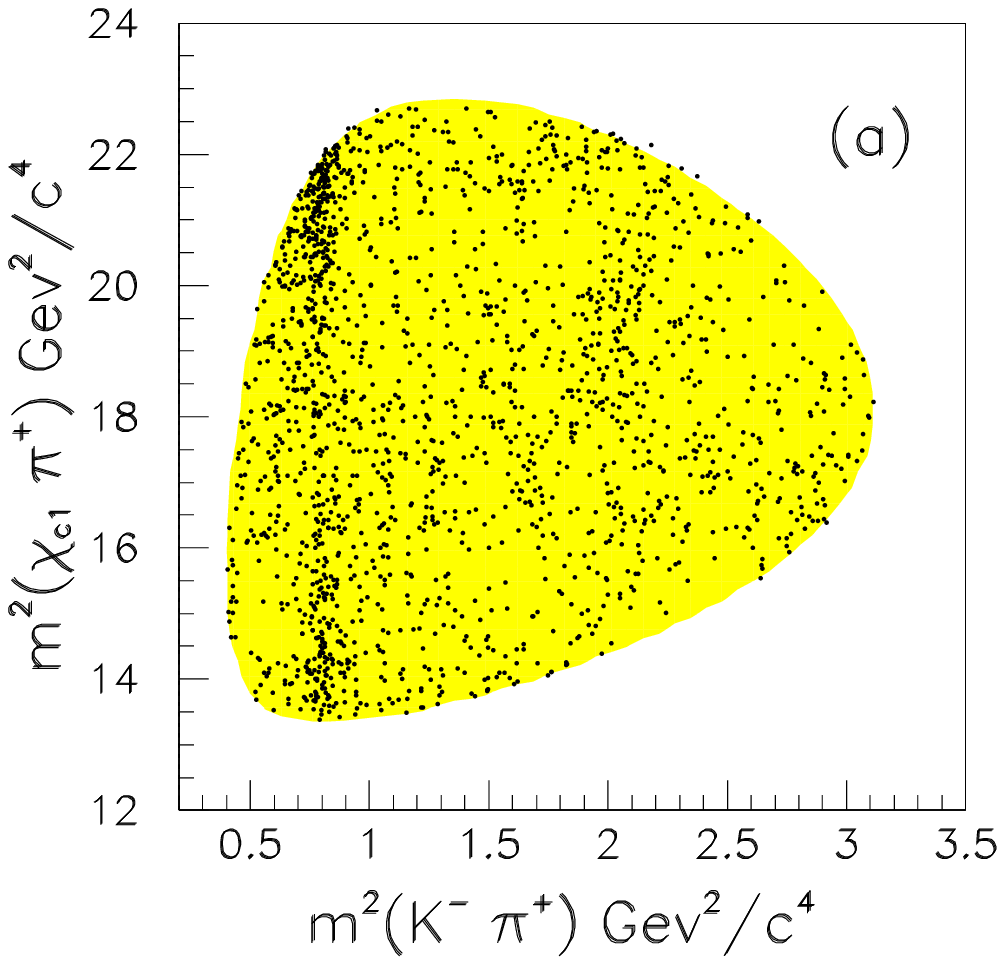}
\includegraphics[width=6cm]{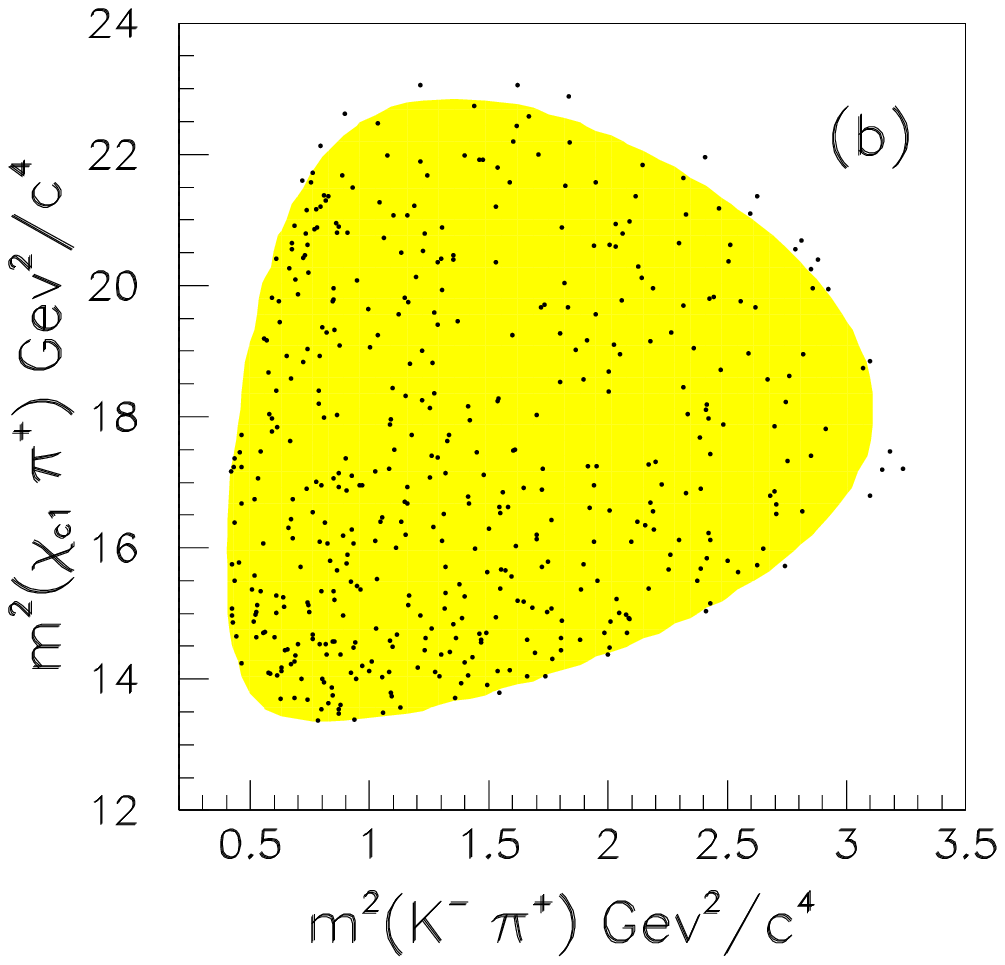}
\caption{Dalitz plot for \Bzchikpi in (a) the signal region and (b) the \DeltaE\ sidebands. The shaded area defines the Dalitz plot boundary.}
\label{fig:fig3}
\end{center}
\end{figure*}

The Dalitz plots for \Bzchikpi events in the signal and sideband regions are shown in Fig.~\ref{fig:fig3}. The shaded
area defines the Dalitz plot boundary; it is obtained from a simple phase space Monte Carlo (MC) simulation~\cite{genbod} of $B$ decays, smeared by the experimental resolution. For the sidebands, events can lie outside the boundary.
\begin{figure*}
\begin{center}
\includegraphics[width=6cm]{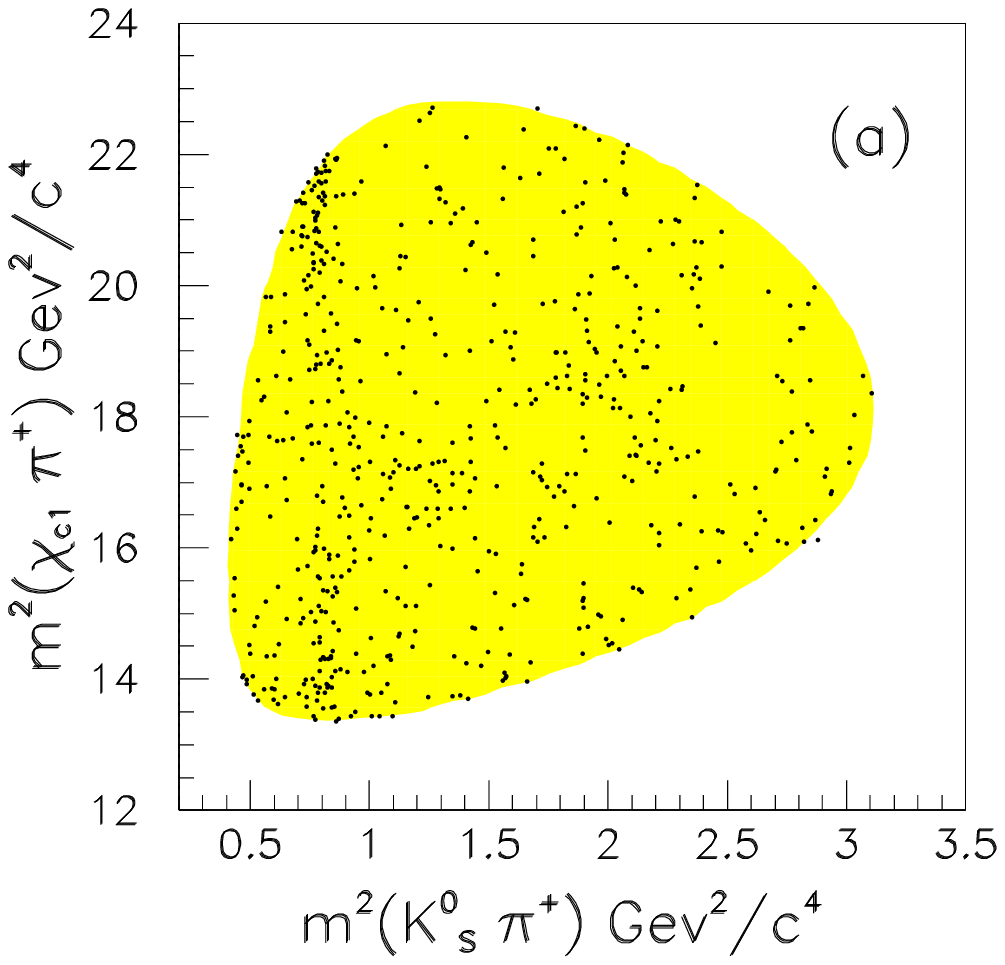}
\includegraphics[width=6cm]{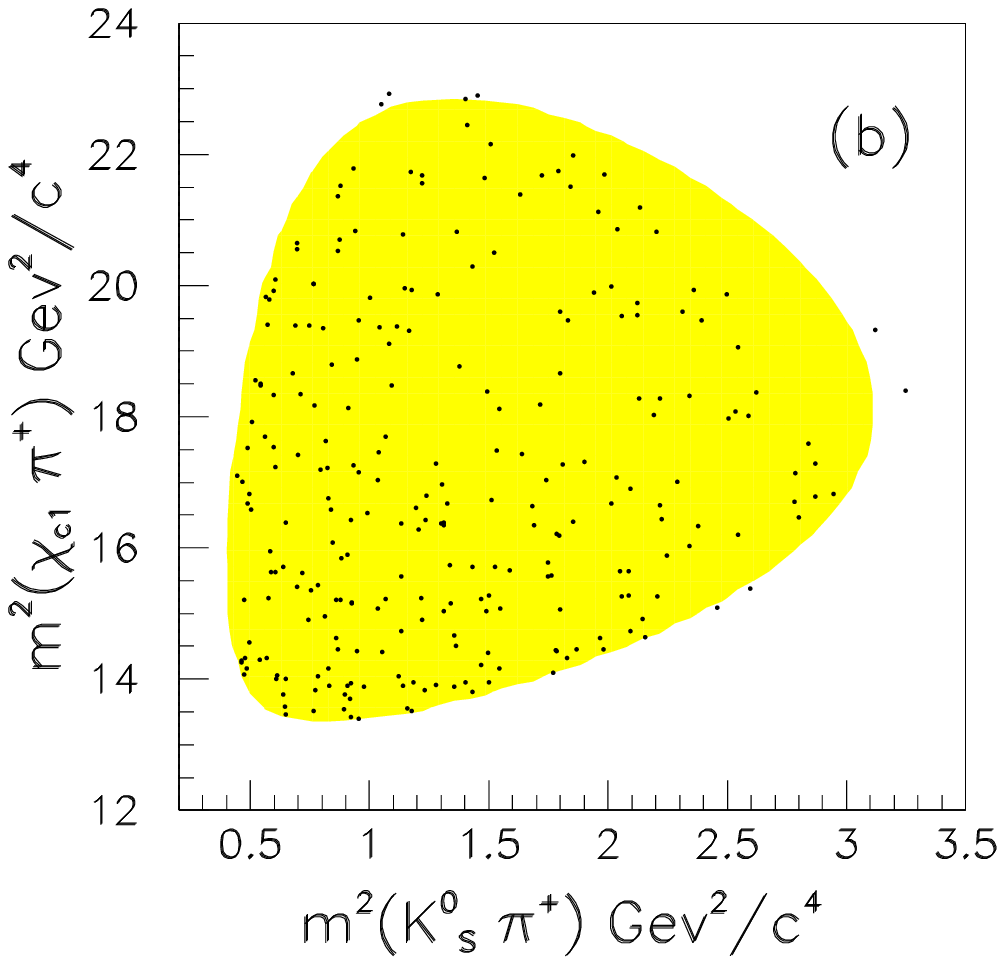}
\caption{Dalitz plot for $\Bp \to \chicone \KS \pip$ in (a) the signal region and (b) the \DeltaE\ sidebands. The shaded area defines the Dalitz plot boundary.}
\label{fig:fig4}
\end{center}
\end{figure*}
We observe a vertical band due to the presence of the $\overline{K^*}(892)^0$ resonance and a weaker band due to the 
$\overline{K^*_2}(1430)^0$
resonance. We do not observe significant accumulation of events in any horizontal band. 

The Dalitz plots for $\Bp \to \chicone \KS \pip$ candidates in the signal and sideband regions are shown in Fig.~\ref{fig:fig4} and show features similar to those in Fig.~\ref{fig:fig3}.

\section{Efficiency}

To compute the efficiency, signal MC events (full-MC) for the different channels have been generated using a detailed detector simulation where $B$ mesons decay uniformly in phase space.
They are reconstructed and analyzed in the same way as real events. We express the efficiency
as a function of $m(K \pi)$ and $\cos \theta$, the normalized dot-product between the $\chicone$ momentum and that of the kaon momentum, both in the
$K \pi$ rest frame. To smooth statistical fluctuations, this efficiency is then parametrized as follows.

\begin{figure}[!htb]
\begin{center}
\includegraphics[width=9cm]{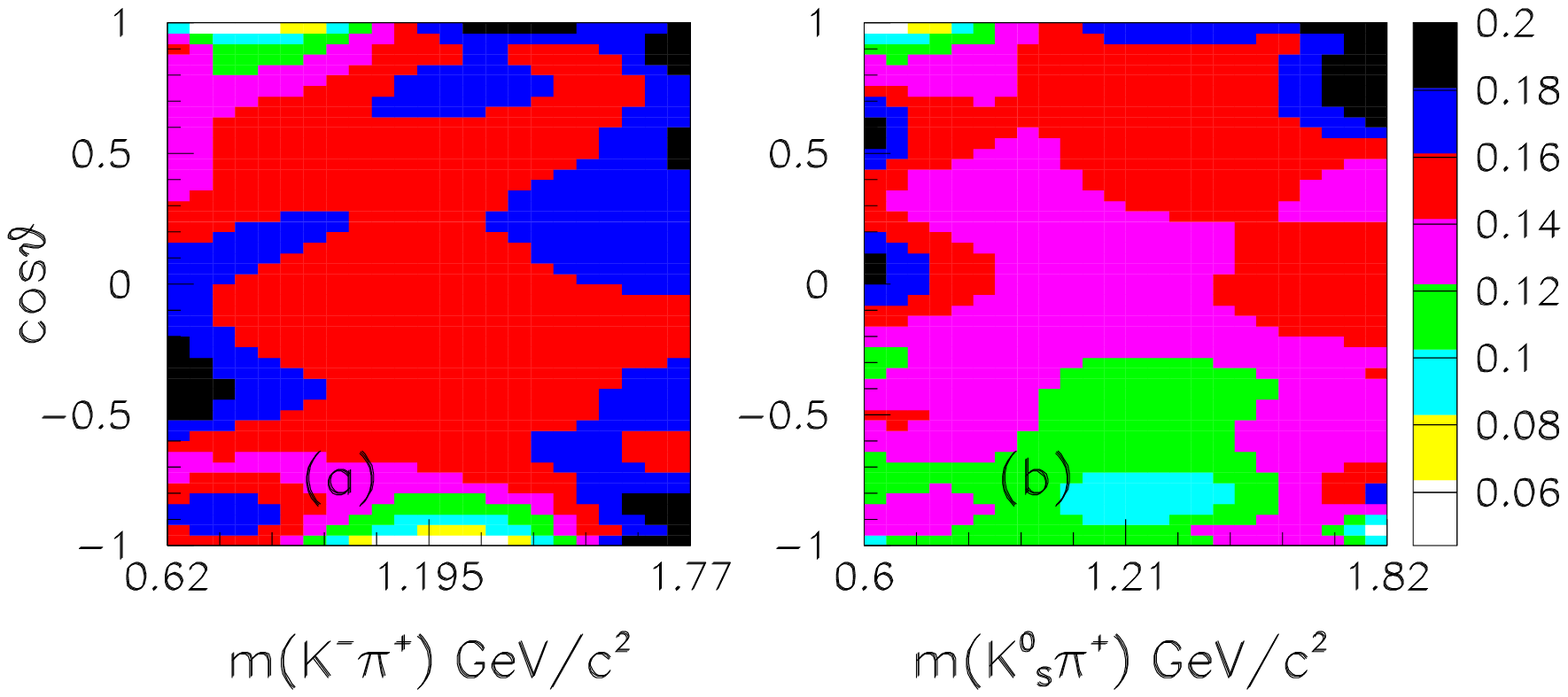}
\caption{Fitted efficiency on the $\cos \theta \  vs. \ m(K \pi)$ plane for (a) \Bzchikpi and 
(b) $\Bp \to \chicone \KS \pip$ summed over the \jpsi decay modes.}
\label{fig:fig5}
\end{center}
\end{figure}

We first fit the efficiency as a function of  $\cos \theta$ in separate 50 \mevcc intervals of $m(K \pi)$, in terms 
of Legendre polynomials up to $L=12$:
\begin{eqnarray}
\epsilon(\cos\theta) = \sum_{L=0}^{12} a_L(m) Y^0_L(\cos\theta).
\end{eqnarray}
For each value of $L$, we fit the $a_L(m)$ as a function of $m(K \pi)$ using a $6^{\rm th}$-order polynomial in $m(K \pi)$.
The resulting fitted efficiency for \Bzb decay is shown in Fig.~\ref{fig:fig5}(a). We observe a significant decrease in
efficiency for $\cos\theta \sim +1$ and $0.72<m(\Km \pip)<0.92$
\gevcc\/, and for $\cos\theta\sim -1$ and $0.97<m(\Km \pip)<1.27$
\gevcc\/. The former is due to the failure to reconstruct
pions with low momentum in the laboratory frame and the latter to a similar
failure for kaons.
A similar effect is observed in Fig.~\ref{fig:fig5}(b) for the the \Bp decay mode.

\begin{figure}[!htb]
\begin{center}
\includegraphics[width=9.5cm]{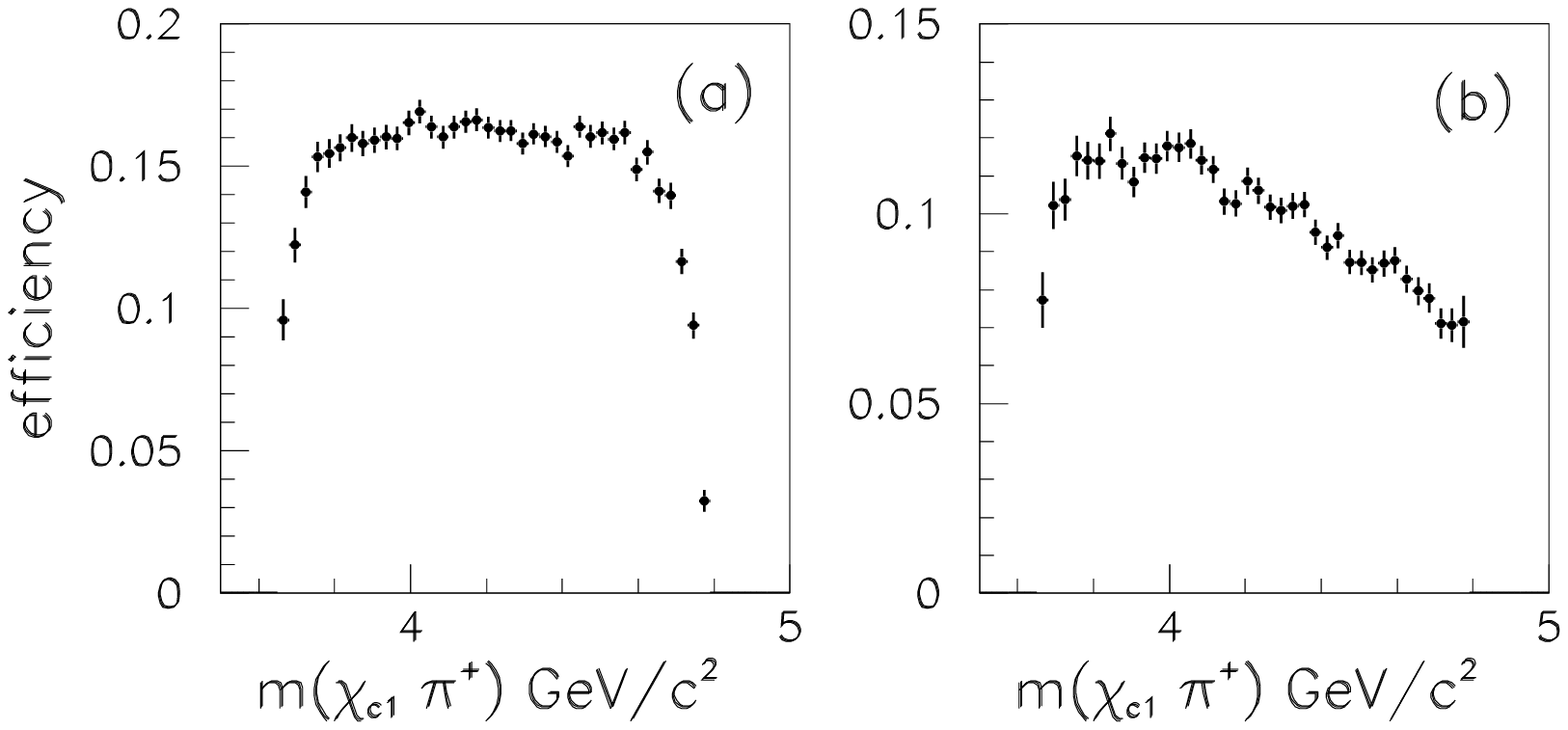}
\caption{Efficiency as a function of $m(\chicone \pip)$ for (a) \Bzchikpi and 
(b) $\Bp \to \chicone \KS \pip$ summed over the \jpsi decay modes.}
\label{fig:fig6}
\end{center}
\end{figure}

In Fig.~\ref{fig:fig6} we plot the efficiency
projection as a function of $m(\chicone \pip)$ for channels (1) and (2), summed over the \jpsi decay modes. 
We observe a loss in efficiency at the edges of the $\chicone \pip$ mass range. However these losses do not affect the 
regions of the reported $Z$ resonances.
Using these fitted functions we obtain efficiency-corrected distributions
by weighting each event by the inverse of the efficiency at its $(m(K \pi), \cos \theta)$ location.

\section{Branching Fractions}
We measure the branching fractions for \Bzchikpi and
$\Bp \to \chicone \KS \pip$ relative to $\Bz \to J/\psi \Kp \pim$ and
$\Bp \to J/\psi \KS \pip$, respectively. In this way several systematic uncertainties,
(namely uncertainties on the number of $B \bar B$ mesons, particle identification, tracking efficiency, data-MC differences, secondary branching fractions) cancel.

To obtain the yields, for each $B$ decay mode we perform new fits to the \DeltaE distributions using the 
full-MC lineshape for the signal and
a linear background. The background-subtracted data are then integrated between $\pm 2.0 \ \sigma_{\DeltaE}$. The correction for efficiency is obtained
as described in Sec. V. A similar procedure is applied to the $\Bzb \to J/\psi \Km \pip$ and $\Bp \to J/\psi \KS \pip$ data.

The branching fraction for $\chicone \to J/\psi \gamma$ from Ref.~\cite{PDG} is $0.344 \pm 0.015$.
Using this value, 
we obtain the following branching fraction ratios:
\begin{eqnarray}
\frac {\BR(\Bzchikpi)}{\BR(\Bzb \to J/\psi \Km \pip)} = 0.474 \pm 0.013 \pm 0.026,
\end{eqnarray}
and
\begin{eqnarray}
\frac {\BR(\Bp \to \chicone K^0 \pip)}{\BR(\Bp \to J/\psi K^0 \pip)} = 0.501 \pm 0.024 \pm 0.028.
\end{eqnarray}
Systematic uncertainties are summarized in Table~\ref{tab:table2} and have been evaluated as follows:
\begin{enumerate}
\item 
We obtain the uncertainty on the background subtraction by modifying the model used to fit the $\Delta E$ distributions. The signal was alternatively
described by the sum of two Gaussian functions and the background was parametrized by a $2^{\rm nd}$-order polynomial.  
\item We compute the uncertainty on the efficiency by making use of the binned efficiency on the $(m(K \pi),\cos \theta)$ plane. In each cell we randomize the generated and reconstructed yields according to Poisson distributions.
Deviations from the fitted efficiencies give the uncertainty on this quantity. 
\item We vary the bin size for the binned efficiency calculation.
\item We include a systematic error due to the uncertainty on the $\chicone \to \jpsi \gamma$ branching fraction~\cite{PDG}.
\item We assign a 1.8 \% uncertainty to the $\gamma$ reconstruction efficiency.
\item We modify the \DeltaE and \mes selection criteria and assign systematic uncertainties based on the variation of the extracted branching fractions.
\end{enumerate}
We note that the systematic uncertainties are dominated by the uncertainty on the $\chicone \to \jpsi \gamma$ branching fraction.
\begin{table}
\caption{Systematic uncertainties (\%) for the $B \to \chicone K \pi$ relative branching fraction measurements.}
\label{tab:table2}
\begin{center}
\vskip -0.2cm
\small{
\begin{tabular}{lcc}
\hline
Contribution & \Bzchikpi  & $\Bp \to \chicone \KS \pip$  \cr
\hline
1. Background subtraction & 1.6 & 1.0 \cr
2. Efficiency & 1.5 & 1.6 \cr
3. Efficiency binning & 1.1 & 1.9 \cr
4. $\chicone$ branching fraction & 4.4 & 4.4 \cr
5. $\gamma$ reconstruction & 1.8 & 1.8 \cr
6. \DeltaE and \mes selections & 1.0 & 1.0\cr
\hline
Total (\%) & 5.4 & 5.5\cr
\hline
\end{tabular}
}
\end{center}
\end{table}

The branching fractions measured in Ref.~\cite{babarz} are:
\begin{eqnarray}
\BR(\Bzb \to J/\psi \Km \pip) = (1.079 \pm 0.011) \times  10^{-3},
\end{eqnarray}
\begin{eqnarray}
\BR(\Bp \to J/\psi K^0 \pip) = (1.101 \pm 0.021) \times  10^{-3},
\end{eqnarray}
where the latter value has been corrected for \KL and $\KS \to \piz \piz$ decays~\cite{PDG}.

Multiplying the ratio in Eq. (4) by the $\Bzb \to J/\psi \Km \pip$ branching fraction in Eq. (6) we obtain
\begin{eqnarray}
\BR(\Bzchikpi) = (5.11 \pm  0.14 \pm 0.28) \times 10^{-4}.
\end{eqnarray}
This may be compared to the Belle measurement~\cite{belle3}: $\BR(\Bzchikpi) = (3.83 \pm 0.10 \pm 0.39) \times 10^{-4}$.

Multiplying the ratio in Eq. (5) by the $\Bp \to J/\psi K^0 \pip$ branching fraction in Eq. (7) we obtain
\begin{eqnarray}
\BR(\Bp \to \chicone K^0 \pip) = (5.52 \pm  0.26 \pm 0.31) \times 10^{-4},
\end{eqnarray}
so that, after all corrections, the branching fractions corresponding to decay modes (1) and (2) are 
the same within uncertainties. 

\section{Fits to the {\boldmath $K\pi$} mass spectra}

We perform binned-$\chi^2$ fits to the background-subtracted and efficiency-corrected $K \pi$ mass spectra in terms of $S$,  $P$, and $D$ wave amplitudes. The fitting function is expressed as: 
\begin{eqnarray}
\lefteqn{\frac{dN}{dm} = N \times } \\
                               & \left[ f_S\frac{G_S(m)}{\int G_S(m) dm}
                               +f_P\frac{G_P(m)}{\int G_P(m) dm} 
                               +f_D\frac{G_D(m)}{\int G_D(m) dm} \right ] \nonumber \ ,
\end{eqnarray}
where $m=m(K\pi)$, the integrals are over the full $m(K\pi)$ range, and the
fractions $f$ are such that
\begin{eqnarray}
 f_S+f_P+f_D=1 \, .
\end{eqnarray}
The $P$- and $D$-wave intensities, $G_P(m)$ and $G_D(m)$, are expressed in
terms of the squared moduli of relativistic Breit-Wigner functions with parameters fixed to the PDG values for $K^*(892)$ and $K^*_2(1430)$ respectively~\cite{PDG}. For the S-wave contribution $G_S(m)$ we make use of the LASS~\cite{lass} parametrization described by Eqs. (11)-(16) of Ref.~\cite{babarz}.

\begin{figure}[!htb]
\begin{center}
\includegraphics[width=9.5cm]{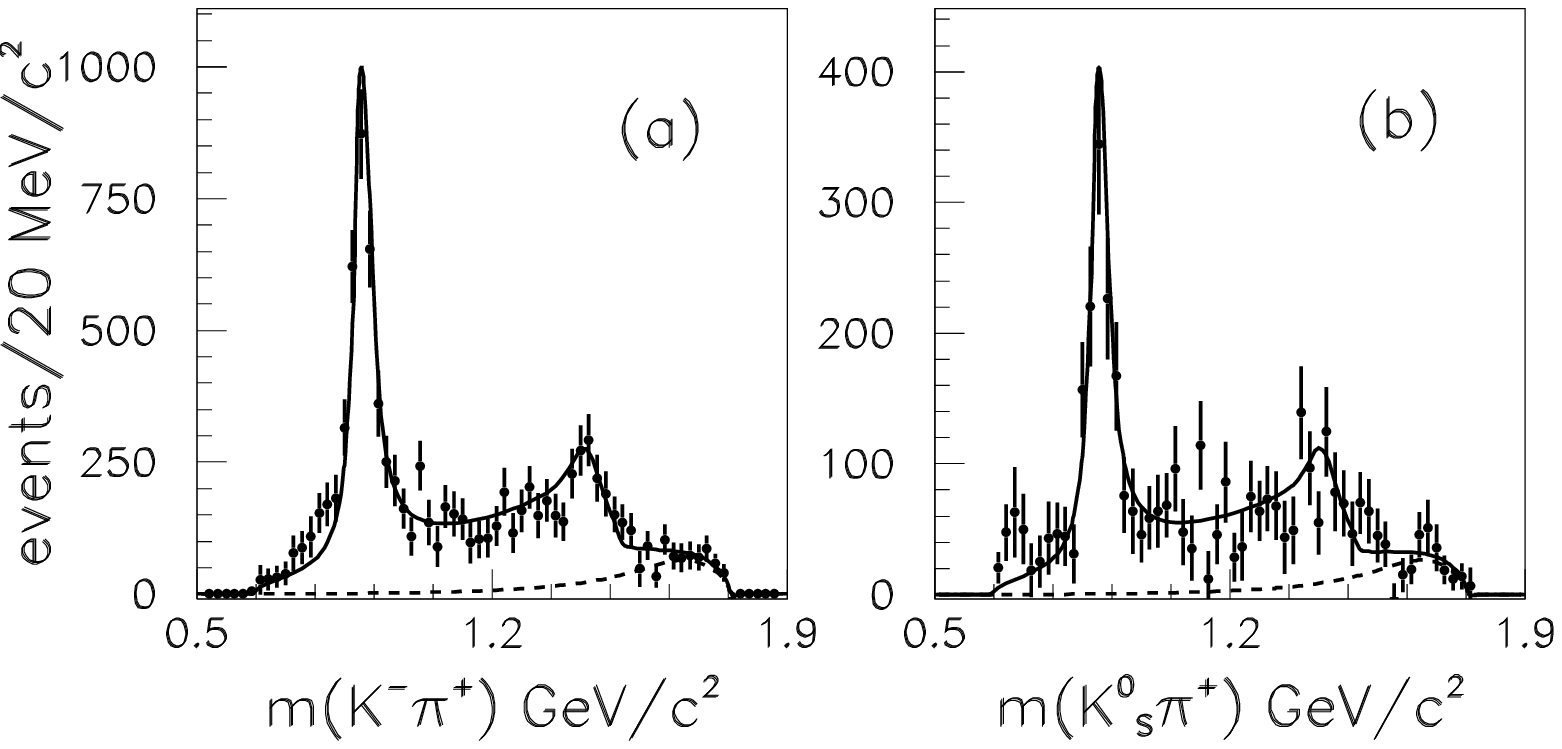}
\caption{Fits to the background-subtracted and efficiency-corrected $K \pi$ mass spectra for (a) \Bzchikpi and
(b) $\Bp \to \chicone \KS \pip$. The $K^*(1680)$ contribution is shown in each figure by the dashed curve.} 
\label{fig:fig7}
\end{center}
\end{figure}

The above model gives a good description of the data for the decays $B \to \jpsi K \pi$~\cite{babarz}.
However, for $B \to \chicone K \pi$
the above resonances do not describe the high mass region of the $K \pi$ mass spectra well. A better fit is obtained  
by including an additional incoherent spin-1 $K^*(1680)$~\cite{PDG} resonance contribution. 
The fit results are shown by the solid curves in Fig.~\ref{fig:fig7}, and the resulting 
intensity contributions are summarized in Table~\ref{tab:table3}.
In Figures~\ref{fig:fig7}(a) and ~\ref{fig:fig7}(b) the contributions due to the $K^*(1680)$ amplitude are shown by the dashed curves. 
The $\chicone K \pi$ decay modes differ from the corresponding $\jpsi K \pi$ and $\psi(2S) K \pi$ decay modes in that the $S$-wave fraction is much larger in the former than in the latter. This
was observed for the $K^*(892)$ region in a previous \babar \ analysis~\cite{babar2}.

\begin{table}
\caption{$S$, $P$, $D$ wave fractions (in \%), and  $\chi^2/{\rm NDF}$ (NDF = Number of Degrees of Freedom)
from the fits to the $K\pi$ mass spectra in \Bzchikpi and $\Bp \to \chicone \KS \pip$. The second $P$-wave entry in the two \chicone channels corresponds to the fraction
of $K^*(1680)$.}
\label{tab:table3}
\begin{center}
\vskip -0.2cm
\begin{tabular}{lcccc}
\hline
Channel & $S$-wave & $P$-wave & $D$-wave & $\chi^2/{\rm NDF}$\cr
\hline
\noalign{\vskip4pt}
\Bzchikpi &  40.4 $\pm$ 2.2 & 37.9 $\pm$ 1.3 & 11.4 $\pm$ 2.0 & 58/54 \cr
        &                 & 10.3 $\pm$ 1.5 &                &  \cr
$\Bp \to \chicone \KS \pip$ &  42.4 $\pm$ 3.5 & 37.1 $\pm$ 3.2 & 10.1 $\pm$ 3.1 & 55/54 \cr
        &                 & 10.4 $\pm$ 2.5 &                &  \cr
\hline
\end{tabular}
\end{center}
\end{table}

\section{The {\boldmath $K\pi$} Legendre polynomial moments}

We compute the efficiency-corrected Legendre polynomial moments $<Y^0_L>$ in each  $K \pi$ mass interval by correcting for efficiency, as explained in Sec. V, and then weighting each event by the $Y^0_L(\cos \theta)$ functions. A similar procedure is performed for the $\Delta E$ sideband events, for
which the distributions are subtracted from those in the signal region.
We observe consistency between the $\Bzb$ and $B^+$ data. Therefore, in the following we combine the \Bzb and $B^+$ distributions. 

\begin{figure}[!htb]
\begin{center}
\includegraphics[width=9.5cm]{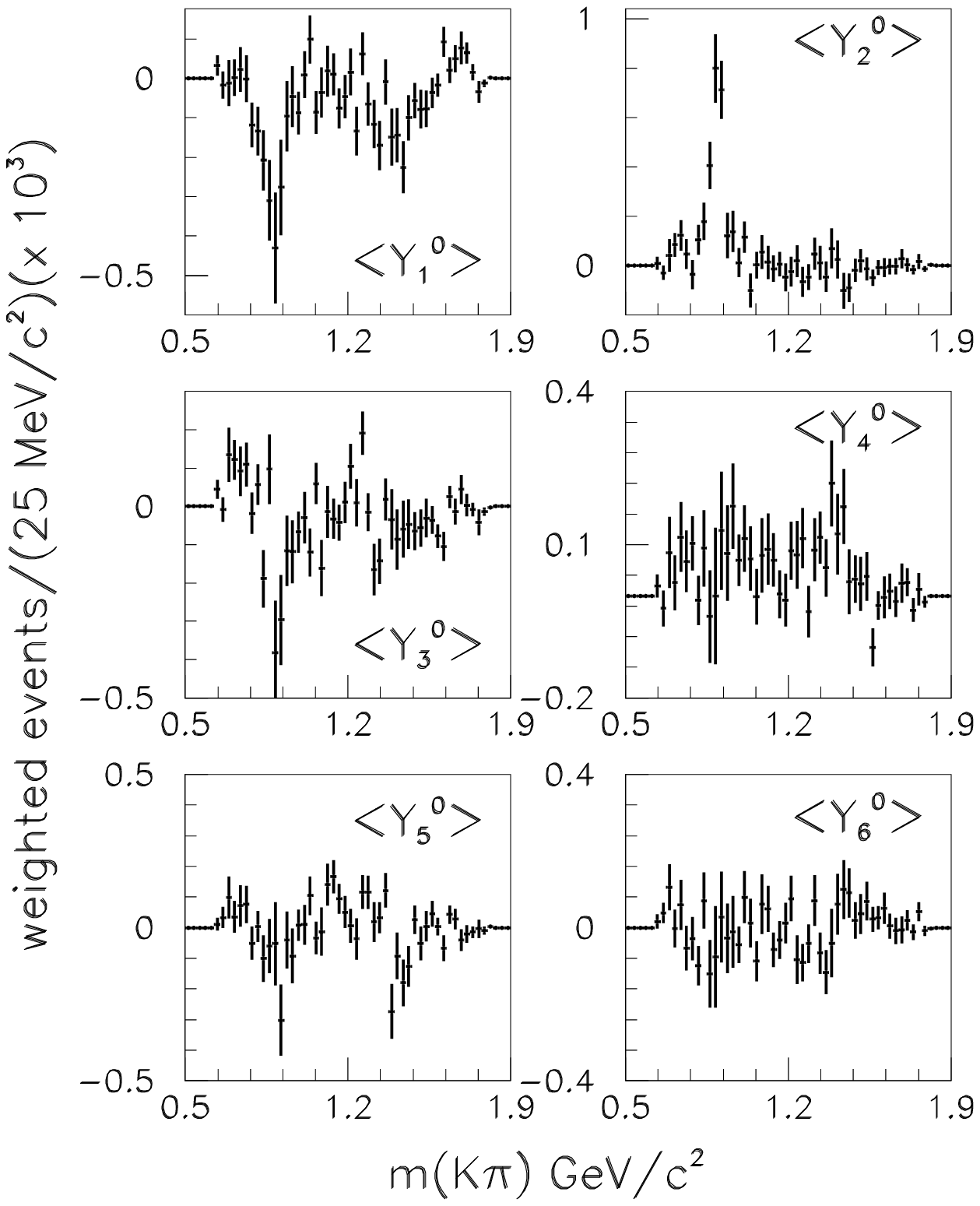}
\caption{Legendre polynomial moments $<Y^0_L>$ for $L=1,...,6$ as functions of $K \pi$ mass for $B \to \chicone K \pi$
after background-subtraction and efficiency-correction.} 
\label{fig:fig8}
\end{center}
\end{figure}

This yields the background-subtracted and efficiency-corrected Legendre polynomial
moments $<Y^0_L>$. They are shown for $L=1,...,6$ in Fig.~\ref{fig:fig8}. We notice that the $<Y^0_6>$ moment is consistent with zero, as are higher moments (not shown).

These moments can be expressed in 
terms of $S$-, $P$- and $D$-wave $K \pi$ amplitudes~\cite{Stephane}. The $P$- and $D$-waves can be present
in three helicity states and, after integration over the decay angles of the \chicone, the relationship between the moments
and the amplitudes is given by Eqs. (26)-(30) of Ref.~\cite{babarz}. 
We notice that, ignoring the presence of resonances in the exotic charmonium channel, the equations involve seven amplitude magnitudes and six
relative phase values, and so they cannot be solved in
each $m(K \pi)$ interval. For this reason, it is not possible to extract the amplitude moduli and relative phase values 
from Dalitz plot analyses of the $\psi K \pi$ or $\chicone K \pi$ final states.

In Fig.~\ref{fig:fig8} we observe the presence of the spin-1
$K^*(890)$ in the $<Y_2^0>$ moment and $S$-$P$ interference in the $<Y_1^0>$ moment. We also observe evidence for the spin-2
$K^*_2(1430)$ resonance in the $<Y_4^0>$ moment. 
There are some similarities between the moments of Fig.~\ref{fig:fig8} and those from 
$B \to \jpsi K \pi$ decays in Ref.~\cite{babarz}. However we also observe a significant structure around 1.7 \gevcc in $<Y_1^0>$ which is absent
in the $B \to \jpsi K \pi$ decays.
We attribute this to the presence of the $K^*_1(1680)$ resonance
produced in $B \to \chicone K \pi$ but absent in $B \to \jpsi K \pi$. The presence of scalar $Z$ resonances should show up especially in
high  $<Y_L^0>$ moments.
 
From the $<Y_L^0>$ we obtain the normalized moments
\begin{equation}
<Y_L^N> = \frac{<Y_L^0>}{n}, 
\end{equation}
where $n$ is the number of events in the given $m(K \pi)$ mass interval.

\section{Monte Carlo Simulations}
We model $B \to \chicone K \pi$ using the resonant structure obtained from the analysis of
the $K \pi$ mass spectra and $K \pi$ Legendre polynomial moments. 
For this purpose we generate a large number of MC events according to the following procedure.
\begin{itemize}
\item{} $B \to \chicone K \pi$ events are generated uniformly in phase space~\cite{genbod}. 
The $B$ mass is generated as a Gaussian lineshape with parameters obtained from a fit to the data. 
\item{} We weight each event by a factor $w_{m(K\pi)}$ derived from the resonant structure in the $K \pi$ system described in Sec. VII (Eq. (10)), and 
displayed in Table~\ref{tab:table3}. 
\item{} We incorporate the measured $K \pi$ angular structure by giving weight $w_{L}$ to each event according to the expression:
\begin{eqnarray}
w_{L}=\sum_{i=0}^{L_{\rm max}}  <Y_i^N> Y^0_i(\cos\theta).
\end{eqnarray}
The moments correspond to the combined data from the decay modes of Eqs. (1) and (2).
The $<Y_i^N>$ are evaluated for the $m(K \pi)$ value by linear interpolation between consecutive $m(K \pi)$ mass intervals.
\item{} The total weight is thus:
\begin{eqnarray}
w = w_{m(K\pi)}\cdot w_{L}
\end{eqnarray}
\end{itemize}

The generated distributions, weighted by the total weight $w$, are then normalized to the number of data events 
obtained after background-subtraction and efficiency-correction.

\begin{figure}[!htb]
\begin{center}
\includegraphics[width=9.5cm]{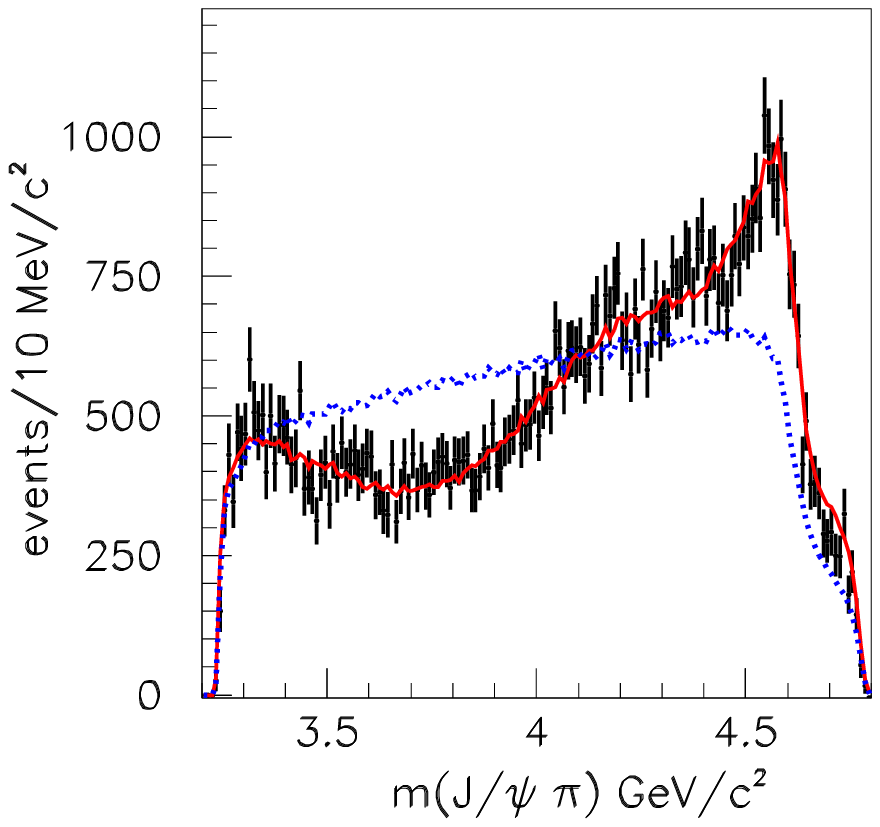}
\caption{Background-subtracted and efficiency-corrected 
$\jpsi \pi$ mass distribution for the $\B \to \jpsi K \pi$ control sample with the superimposed curves resulting from the MC
simulation described in the text.
The solid curve is obtained using the total weight $w$ obtained with $L_{\rm max}=5$, the dotted curve by
omitting the angular-dependence factor $w_{L}$.} 
\label{fig:fig9}
\end{center}
\end{figure}

We first test the method using as control sample the combined data from $\Bzb \to \jpsi \Km \pip$ and $\Bp \to \jpsi \KS \pip$, where no resonant structure is observed in the $\jpsi \pi$ 
mass distributions~\cite{babarz}. In this case we generate $\B \to \jpsi K \pi$ events and use the $K \pi$ resonant structure and Legendre polynomial 
information from the same channels.  
We compare the MC simulation to the $\jpsi \pip$ mass projection from data in Fig.~\ref{fig:fig9}. We obtain $\chi^2/{\rm NDF}=223,162,180/152$ for $L_{\rm max}=4,5,6$
respectively. We conclude that $L_{\rm max}=5$ gives the best description of the data.

\begin{table}
\caption{The value of $\chi^2/{\rm NDF}$ for different MC-data comparisons; ``$Y^N_L$'' indicates the channel used
to obtain the normalized moments. The ``mixed'' algorithm is explained in the text. The definition of window is given in Sec. X.}
\label{tab:table4}
\begin{center}
\vskip -0.2cm
\begin{tabular}{lccc}
\hline
Channel & $Y^N_L$  &  $L_{\rm max}$ & $\chi^2/{\rm NDF}$ \cr
\hline
$B \to J/\psi K \pi$ & $B \to J/\psi K \pi$ & 5 &  162/152 \cr
$B \to \chicone K \pi$ & $B \to \chicone K \pi$ & 5 & 46/58 \cr
\hline
$B \to \chicone K \pi$ & $B \to \chicone K \pi$ & ``mixed'' & 63/58 \cr
\hline
$B \to \chicone K \pi$ window & $B \to \chicone K \pi$ & 5 & 45/47 \cr
$B \to \chicone K \pi$ window & $B \to \chicone K \pi$ & ``mixed'' & 56/47 \cr
\hline
\end{tabular}
\end{center}
\end{table}

\begin{figure}[!htb]
\begin{center}
\includegraphics[width=9.5cm]{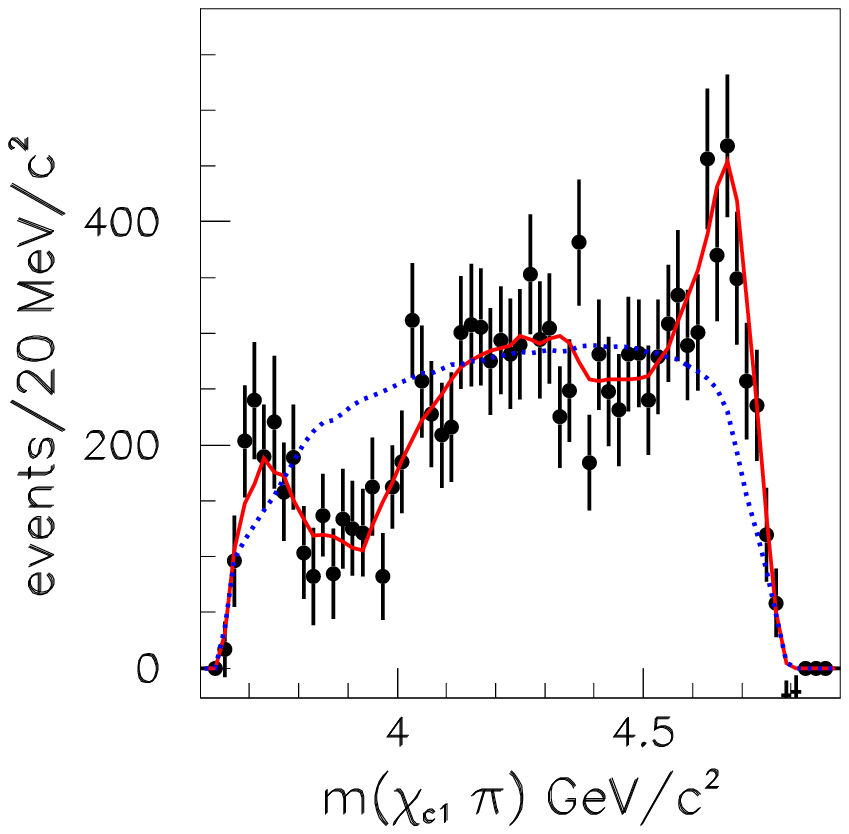}
\caption{Background-subtracted and efficiency-corrected $\chicone \pi$ 
mass distribution from $B \to \chicone K \pi$. The solid curve results from the MC
simulation described in the text, which uses of the moments from the same channels. The dotted curve
shows the result of the simulation when the $w_L$ weight is removed.}
\label{fig:fig10}
\end{center}
\end{figure}

We now perform a similar MC simulation for $B \to \chicone K \pi$ using moments from the same channels.
We obtain $\chi^2/{\rm NDF}=53,46,49/58$ for $L_{\rm max}=4,5,6$
respectively. The result of the simulation with  $L_{\rm max}=5$ is superimposed on the data in Fig.~\ref{fig:fig10}, and the corresponding
$\chi^2/{\rm NDF}$ is given in Table~\ref{tab:table4}. The excellent description of the data
indicates that the angular information from the $K \pi$ channel with $L_{\rm max}=5$ 
is able to account for the structures observed in 
the $\chicone \pi$ projection. This indicates the absence of significant structure in the exotic $\chicone \pip$ channel.

\begin{figure}[!htb]
\begin{center}
\includegraphics[width=9.5cm]{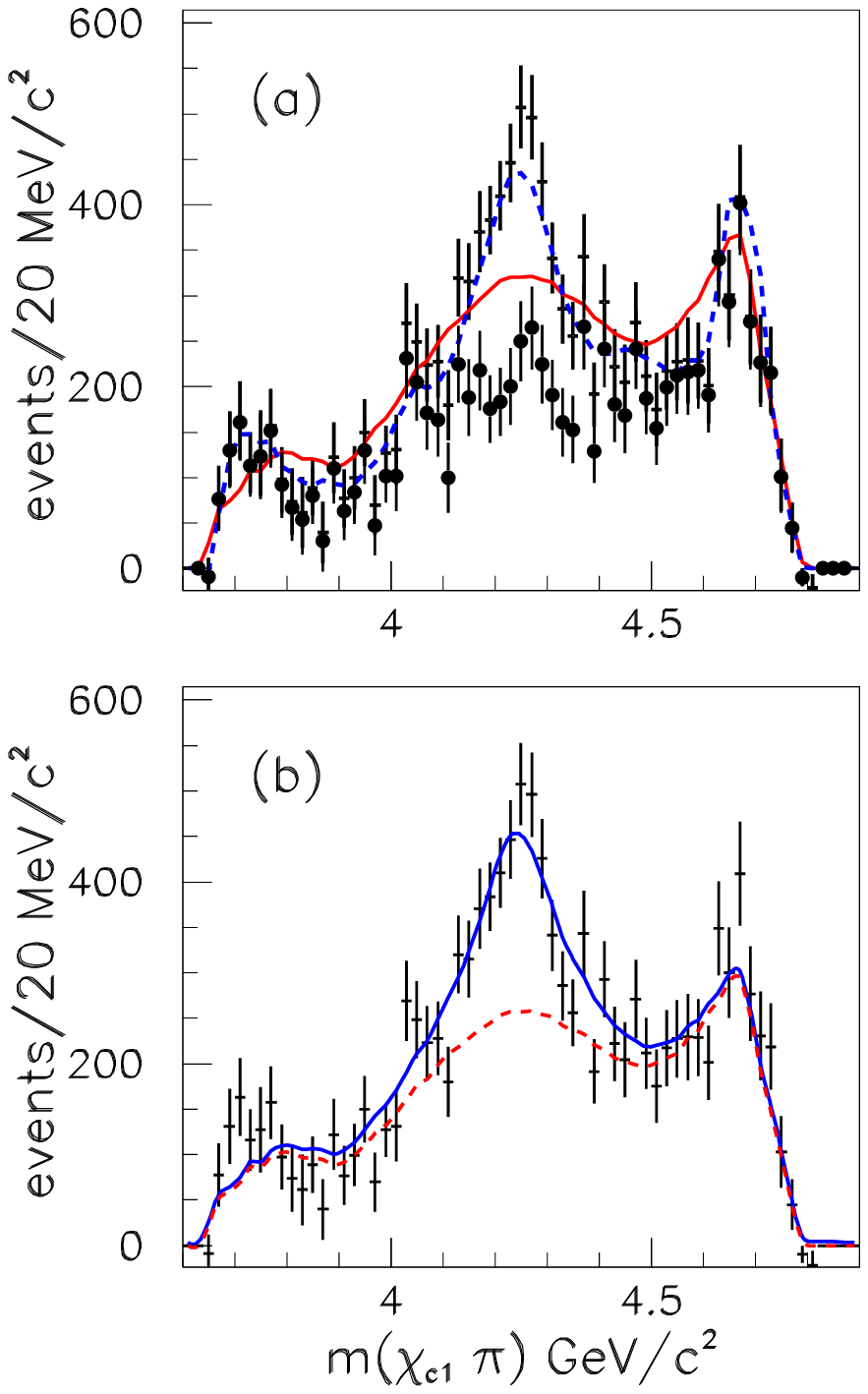}
\caption{Background-subtracted and efficiency-corrected  $\chicone \pi$ 
mass distribution from $B \to \chicone K \pi$ which includes a simulated $\Zb$ (vertical crosses). In (a) the distribution with solid dots represents
the $\Bzb \to \chicone K^- \pi^+$ data component. The continuous curve is the result from the ``mixed''
simulation described in the text and obtained from the MC simulation. The dashed curve shows a simulation with $L_{\rm max}=15$. 
(b) Result from the fit described in the text, which incorporates a Breit-Wigner lineshape
 describing the \Zb. The dashed curve represents the background model from the ``mixed'' simulation.} 
\label{fig:fig11}
\end{center}
\end{figure}

We perform a MC simulation where, to the data from \Bzchikpi, we add an arbitrary fraction ($\approx$ 25 \%) of events which 
include a $\Zb$ resonance
 decaying to $\chicone \pi$.  These $\Zb$ events are obtained from phase-space MC  
$\Bzb \to \chicone K^- \pi^+$ events 
weighted by a simple Breit-Wigner. We then compute Legendre polynomial moments for the total sample and use them to predict the $\chicone \pi$ mass distribution as described above.
The $\chicone \pi$ mass spectrum for these events is shown in Fig.~\ref{fig:fig11}(a).
We obtain $\chi^2/{\rm NDF}=103, 91, 88/58$ for $L_{\rm max}=4,5,6$ respectively. Therefore, in the presence of a $\Zb$ resonance, it is not possible
to obtain a good description of the $\chicone \pi$ mass distribution using $L_{\rm max}=5$. We then increase the value of $L_{\rm max}$ and 
obtain a good description of this MC simulation with $L_{\rm max}=15$, as shown by the dashed
curve in Fig.~\ref{fig:fig11}(a) ($\chi^2/{\rm NDF}=57/58$).

We next test a ``mixed'' simulation where we use $L_{\rm max}=3$ up to a $K \pi$ mass of 1.2 \gevcc and $L_{\rm max}=4$ for the rest of the 
events. This choice is justified by the presence of spin 0 and 1 resonances mostly in the low $K \pi$ mass region, while the $K^*_2(1430)$
contributes for $m(K \pi) > 1.2$ \gevcc. This simulation gives a satisfactory description of the $B \to \chicone K \pi$ 
data with $\chi^2/{\rm NDF}=63/58$ 
but gives a bad description of the MC sample of Fig.~\ref{fig:fig11}(a), yielding $\chi^2/{\rm NDF}=140/58$.

We now fit the MC sample including a simple Breit-Wigner (with the width fixed to the simulated value) to describe the $\Zb$ 
(Fig.~\ref{fig:fig11}(b)). We obtain the solid curve, which has $\chi^2/{\rm NDF}=75/56$, a $\Zb$ mass consistent with the generated value, and a yield consistent with the generated one. The dashed curve represents the background model from the ``mixed'' simulation. The MC test therefore validates the use of this background model for a quantitative evaluation of the upper limits described in Sec. X. 

The data-MC comparisons for the different simulations are summarized
in Table~\ref{tab:table4}. 

\section{Search for {\boldmath $\Za$} and {\boldmath $\Zb$}} 

We have shown, in the previous sections, that in the absence of $Z$ resonances, the simulation with $L_{\rm max}=5$ gives a good
description of the $\B \to \jpsi K \pi$ and $B \to \chicone K \pi$ data.
We now test the possible presence of the \Za and \Zb resonances in $B \to \chicone K \pi$ decay.
Therefore we adopt the minimum $L_{\rm max}$ configuration ("mixed") described in Sec. IX and investigate whether something else
is needed by the data.

For this purpose we perform binned $\chi^2$ fits to 
the $\chicone \pip$ mass spectrum. 
In these fits the normalization of the background component is determined by the fit.
We observe that this background model predicts an enhancement in the mass region of the $Z$ resonances.
We then add, for the signal, 
relativistic spin-0 Breit-Wigner functions with parameters fixed to the Belle values for the signals~\cite{belle3}.
We compute statistical significance using the fitted fraction divided by its uncertainty.

\begin{figure*}[!htb]
\begin{center}
\includegraphics[width=14cm]{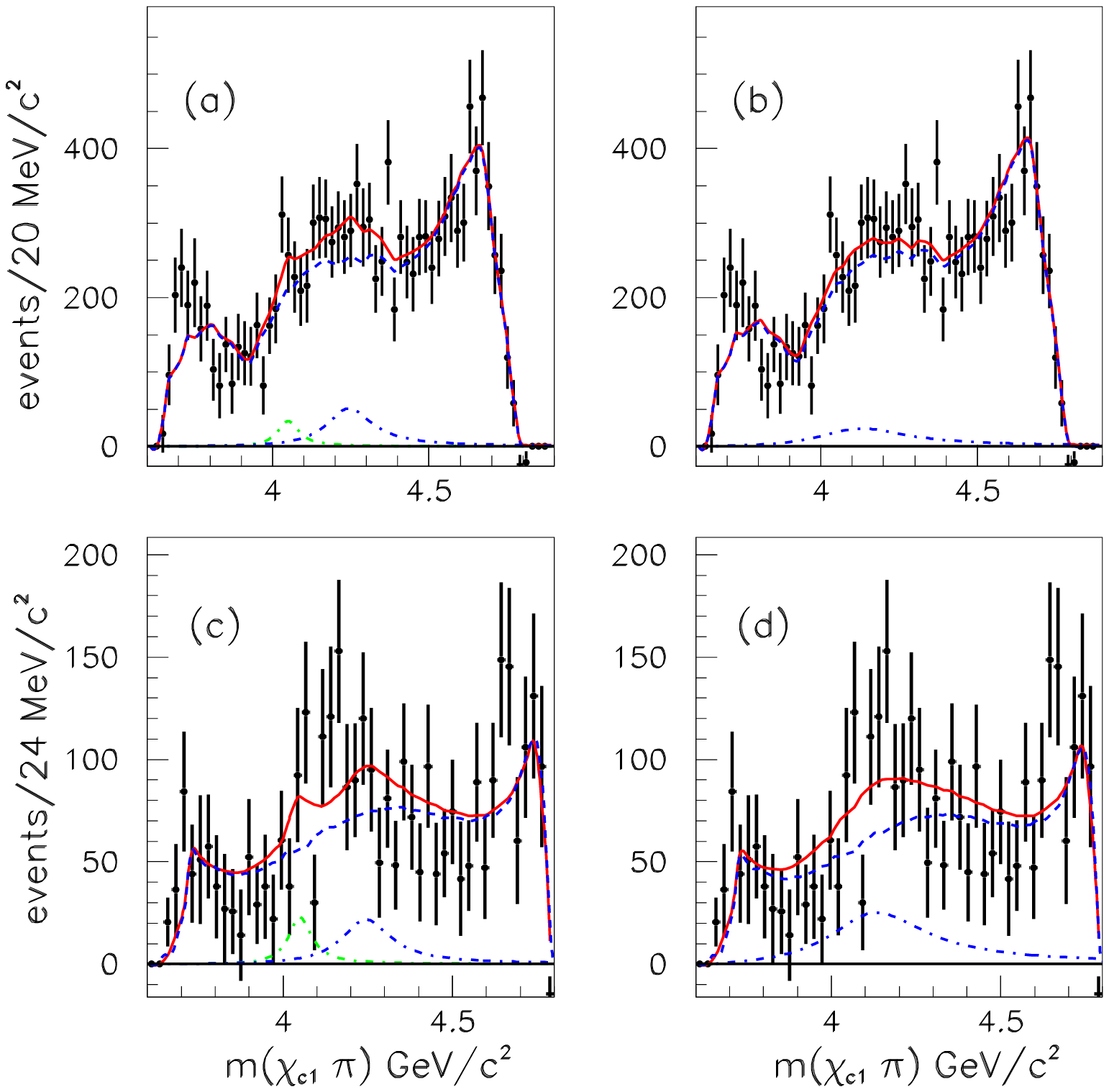}
\caption{(a),(b) Background-subtracted and efficiency-corrected $\chicone \pi$ 
mass distribution for $B \to \chicone K \pi$. 
(a) Fit with \Za and \Zb resonances.
(b) Fit with only the $Z(4150)^+$ resonance.
(c),(d) Efficiency-corrected and background-subtracted $\chicone \pi$ 
mass distribution for $B \to \chicone K \pi$ in the $K \pi$ mass region $1.0 < m^2(K \pi) < 1.75$ \gevcccc.
(c) Fit with \Za and \Zb resonances.
(d) Fit with only the $Z(4150)^+$ resonance. In each fit the dashed curve shows the prediction from the ``mixed'' $B \to \chicone K \pi$ simulation explained in the text. The dot-dashed curves indicate the fitted resonant contributions.}
\label{fig:fig12}
\end{center}
\end{figure*}
We first perform fits to the total mass spectrum.
\begin{itemize}
\item{} Fit a) is shown in Fig.~\ref{fig:fig12}(a), and includes both \Za and \Zb resonances.
\item{} Fit b) is shown in Fig.~\ref{fig:fig12}(b), and includes a single broad $Z(4150)^+$ resonance.
\end{itemize}
In both cases the fits give fractional contributions consistent with zero for the $Z$ resonances.

We next fit the $\chicone \pi$ mass spectrum in the Dalitz plot region $1.0 \le m^2(K \pi)<1.75 \ \gevcccc$ in order to make a
direct comparison to the Belle results~\cite{belle3}. 
Figures~\ref{fig:fig12}(c),(d) show the $\chicone \pi$ mass spectrum for this mass region 
(labeled as ``window'' in Table~\ref{tab:table5}) where the Belle data show the maximum of the reported resonance activity. 
This sample accounts for 25 \% of our total data sample. Table~\ref{tab:table4} gives the corresponding $\chi^2/{\rm NDF}$ values for the MC simulations described in Sec. IX, in this mass window.
\begin{itemize}
\item{} Fit c) is shown in Fig.~\ref{fig:fig12}(c), and includes both \Za and \Zb resonances. 
\item{} Fit d) is shown in  Fig.~\ref{fig:fig12}(d), and includes a single broad $Z(4150)^+$ resonance.
\end{itemize}
In each case the fit gives a $Z$ resonance contribution consistent with zero.
\begin{table}
\caption{Results of the fits to the  $\chicone \pi$ mass spectra. $N_{\sigma}$ and Fraction give, for each fit, the significance
and the fractional contribution of the $Z$ resonances.}
\label{tab:table5}
\begin{center}
\vskip -0.2cm
\begin{tabular}{lcccc}
\hline
 Data & Resonance & $N_{\sigma}$ & Fraction (\%) & $\chi^2/NDF$\cr
\hline
a) Total & \Za & 1.1 & 1.6 $\pm$ 1.4 & 57/57 \cr
                      & \Zb & 2.0 & 4.8 $\pm$ 2.4  & \cr
b) Total  & $Z(4150)^+$ & 1.1 & 4.0 $\pm$ 3.8 & 61/58 \cr
\hline
c) Window  & \Za & 1.2 & 3.5 $\pm$ 3.0 & 53/46 \cr
                      & \Zb & 1.3 & 6.7 $\pm$ 5.1 \cr
d) Window  & $Z(4150)^+$ & 1.7 & 13.7 $\pm$ 8.0 & 53/47 \cr
\hline
\end{tabular}
\end{center}
\end{table}

The results of the fits are summarized in  Table~\ref{tab:table5}, and in every case the 
yield significance does not exceed 2$\sigma$.
Similar results are obtained when the resonance parameters are varied within their statistical errors.

We compute upper limits integrating the region of positive branching fraction values for a Gaussian function having the above mean and $\sigma$ values, and
obtain the following  90\% C.L. limits for the \Za and \Zb resonances:
\begin{eqnarray}
\BR(\Bzb \to \Za  K^-)\times\BR(\Za \to \chicone \pip)\\
<1.8 \times 10^{-5},\nonumber \\
\BR(\Bzb \to \Zb  K^-)\times\BR(\Zb \to \chicone \pip)\\
<4.0 \times 10^{-5},\nonumber\\
\BR(\Bzb \to Z^+  K^-)\times\BR(Z^+ \to \chicone \pip)\\
<4.7 \times 10^{-5}.\nonumber
\end{eqnarray}
Systematic uncertainties related to the $Z$ parameters have been ignored since they give negligible contributions.
The corresponding values for \Bp decay are $\approx$8\% larger (see Eqs. (8) and (9)).

Our measurements can be compared to the Belle results~\cite{belle3}:
\begin{eqnarray}
\BR(\Bzb \to \Za  K^-)\times\BR(\Za \to \chicone \pip)\\
= 3.0^{+1.5 \ +3.7}_{-0.8 \ -1.6} \times 10^{-5},\nonumber \\
\BR(\Bzb \to \Zb  K^-)\times\BR(\Zb \to \chicone \pip)\\
= 4.0^{+2.3\ +19.7}_{-0.9 \ \ -0.5} \times 10^{-5},\nonumber\\
\end{eqnarray}
Given the large uncertainties, these branching fraction values are compatible with our upper-limit estimates. 

\section{Conclusions}
We use 429 fb$^{-1}$ of data from the \babar\ experiment at SLAC to search for the \Za and \Zb states decaying to $\chicone \pip$ in the
decays \Bzchikpi and $\Bp \to \chicone \KS \pip$, where 
$\chicone \to \jpsi \gamma$ .

We measure the following branching fractions for the decays \Bzchikpi and  $\Bp \to \chicone K^0 \pip$:

$$\BR(\Bzchikpi) = (5.11 \pm  0.14 \pm 0.28) \times 10^{-4},$$
and
$$\BR(\Bp \to \chicone K^0 \pip) = (5.52 \pm  0.26 \pm 0.31) \times 10^{-4}.$$

In our search for the $Z$ states, we first attempt
to describe the data assuming that all resonant activity is 
concentrated in the $K \pi$ system. We use the decay $B \to \jpsi K \pi$ as a control sample, since no resonant structure 
has been observed in the $\jpsi \pi$ mass
spectrum. In this case a good description of the data is obtained by a MC simulation which makes use of the known
resonant structure in the $K \pi$ mass spectrum together with a Legendre-polynomial description
of the angular structure as a function of $K \pi$ mass.

The same procedure is then applied to our data on the decays $B \to \chicone K \pi$ and a good description of the
$\chicone \pi$ mass distribution is obtained. 
This indicates that no significant resonant structure
is present in the $\chicone \pi$ mass spectrum, as observed for the $\jpsi \pi$ mass distribution~\cite{babarz}. 
We also observe that this background model predicts an enhancement in the mass region of the $Z$ resonances.
We then report 90\% C.L. upper limits on possible $\Bzb \to Z^+ K^-$ decays. 

In conclusion, we find that it is possible to obtain a good description of our data without the need for additional resonances in the 
$\chicone \pi$ system. 
\section{Acknowledgements}
We are grateful for the 
extraordinary contributions of our \pep2\ colleagues in
achieving the excellent luminosity and machine conditions
that have made this work possible.
The success of this project also relies critically on the 
expertise and dedication of the computing organizations that 
support \babar.
The collaborating institutions wish to thank 
SLAC for its support and the kind hospitality extended to them. 
This work is supported by the
US Department of Energy
and National Science Foundation, the
Natural Sciences and Engineering Research Council (Canada),
the Commissariat \`a l'Energie Atomique and
Institut National de Physique Nucl\'eaire et de Physique des Particules
(France), the
Bundesministerium f\"ur Bildung und Forschung and
Deutsche Forschungsgemeinschaft
(Germany), the
Istituto Nazionale di Fisica Nucleare (Italy),
the Foundation for Fundamental Research on Matter (The Netherlands),
the Research Council of Norway, the
Ministry of Education and Science of the Russian Federation, 
Ministerio de Ciencia e Innovaci\'on (Spain), and the
Science and Technology Facilities Council (United Kingdom).
Individuals have received support from 
the Marie-Curie IEF program (European Union), the A. P. Sloan Foundation (USA) 
and the Binational Science Foundation (USA-Israel).


\begin{thebibliography}{99}
\bibitem{conj}
The use of charge conjugate reactions is implied throughout this work.
\bibitem{brem}
The \jpsi vertex position is used to define the three-momentum
vector of the photon, and this is then added to the three-momentum
of the associated $e^+$ or $e^-$ at this vertex. The electron mass is
then assigned to this modified three-momentum vector, and the
resulting four-momentum vector is used in calculating $m(\ee)$.
\end{thebibliography}
\end{document}